\documentclass{elsart4-1}

\usepackage{graphicx}
\usepackage{epsfig}

\usepackage{amssymb}

\usepackage[english,francais]{babel}
\usepackage[numbers]{natbib}
\usepackage{subfigure}

\newtheorem{e-proposition}[theorem]{Proposition}

\newtheorem{e-definition}[theorem]{Definition\rm}

\renewcommand{\theequation}{\arabic{equation}}
\newcommand{\g}{$\gamma$}

\def\og{\leavevmode\raise.3ex\hbox{$\scriptscriptstyle\langle\!\langle$~}}
\def\fg{\leavevmode\raise.3ex\hbox{~$\!\scriptscriptstyle\,\rangle\!\rangle$}}

\begin{document}

\centerline{Astrophysics}
\begin{frontmatter}



\selectlanguage{english}
\title{Starburst galaxies as seen by \g-ray telescopes}


\selectlanguage{english}
\author{Stefan Ohm}
\ead{stefan.ohm@desy.de}

\address{Deutsches Elektronen Synchrotron DESY, 15738 Zeuthen,
  Germany}

\medskip
\begin{center}
{\small Received *****; accepted after revision +++++}
\end{center}

\begin{abstract}
  Starburst galaxies have a highly increased star-formation rate
  compared to regular galaxies and inject huge amounts of kinetic
  power into the interstellar medium via supersonic stellar winds, and
  supernova explosions. Supernova remnants, which are considered to be
  the main source of cosmic rays (CRs), form an additional,
  significant energy and pressure component and might influence the
  star-formation process in a major way. Observations of starburst
  galaxies at \g-ray energies gives us the unique opportunity to study
  non-thermal phenomena associated with hadronic CRs and their
  relation to the star-formation process. In this work, recent
  observations of starburst galaxies with space and ground-based
  \g-ray telescopes are being reviewed and the current state of
  theoretical work on the \g-ray emission is discussed. A special
  emphasis is put on the prospects of the next-generation Cherenkov
  Telescope Array for the study of starburst galaxies in particular
  and star-forming galaxies in general.

{\it To cite this article: S. Ohm, C. R. Physique XX (2015).}
\vskip 0.5\baselineskip

\noindent{\small{\it Keywords~:} Cosmic rays; gamma rays; star
  clusters; starburst galaxies; ultra luminous infrared galaxies
  \vskip 0.5\baselineskip}

\selectlanguage{francais}
\noindent{\bf R\'esum\'e}
\vskip 0.5\baselineskip
\noindent

Les galaxies \`a flamb\'ees d'\'etoiles se caract\'erisent par un taux
de formation d'\'etoiles beaucoup plus \'elev\'e que ceux des galaxies
ordinaires. Les vents stellaires supersoniques et les explosions de
supernov\ae~qui s'y produisent injectent dans le milieu interstellaire
une \'energie cin\'etique consid\'erable.  De plus, alors que les
vestiges de supernov\ae~sont consid\'er\'es comme les sources
principales de rayons cosmiques, ces derniers augmentent de mani\`ere
significative la pression et la densit\'e d'\'energie du milieu, au
point d'influencer fortement le processus de formation
d'\'etoiles. L'observation de galaxies \`a flamb\'ees d'\'etoiles en
astronomie gamma est un moyen unique pour \'etudier les ph\'enom\`enes
non thermiques dus \`a des protons et noyaux cosmiques, et leur r\^ole
dans le processus de formation d'\'etoiles. Cet article passe en revue
les observations r\'ecentes de galaxies \`a flamb\'ees d'\'etoiles
avec des t\'elescopes \`a rayons gamma dans l'espace et \`a partir du
sol. Il discute aussi les interpr\'etations th\'eoriques actuelles de
l'\'emission gamma observ\'ee. Enfin, un accent particulier est mis
sur l'impact des t\'elescopes \`a effet Tcherenkov atmosph\'erique de
la prochaine g\'en\'eration sur l'\'etude des galaxies \`a flamb\'ee
d'\'etoiles en particulier et, plus g\'en\'eralement, sur la formation
d'\'etoiles dans les galaxies.

{\it Pour citer cet article~: S. Ohm, C. R. Physique XX (2015).}

\vskip 0.5\baselineskip
\noindent
\vskip 0.5\baselineskip
\noindent{\small{\it Mots­cl\'es~:} Rayons cosmiques~; rayons gamma~; amas d'\'etoiles~; galaxies \`a flamb\'ee d'\'etoiles~; galaxies infrarouges ultra­lumineuses}

\end{abstract}
\end{frontmatter}


\selectlanguage{english}

\section{Introduction}
\label{sec:intro}

The term ``starburst'' is often used to describe regions of greatly
enhanced star-formation within galaxies or to characterise entire
galaxies. The starburst phenomenon covers a wide range of physical
scales from blue compact starburst-galaxies, circumnuclear rings in
local barred galaxies to luminous (LIRGs) and ultraluminous infrared
galaxies (ULIRGs). The star-formation rate (SFR) in starbursts is
considered to be out of equilibrium, with gas-consumption timescales
of 1\,Gyr or shorter, and typical timescales for the starburst episode
of a few 100\,Myrs (see e.g. \cite{Kennicutt2012, Krumholz2014} and
references therein).

Cosmic rays (CRs) are considered to be an important star-formation
regulator since they penetrate deep into molecular cloud cores: the
seeds of protostars. In fact they penetrate much deeper into clouds
than UV radiation, which is effectively shielded from the most dense
cores (see e.g. \cite{Caselli1998, Dalgarno2006}). CRs initiate
complex chemical reactions and are the main driver for gas-phase
chemistry in the interstellar medium (ISM) \cite{Indriolo2012,
  Indriolo2013}. In regions where CR ionisation rates are very high
(e.g starbursts and ULIRGs), CRs might even influence the initial
conditions of star formation by preventing low-mass star formation,
which leads to a top-heavy initial mass function \cite{Socrates2008,
  Papadopoulos2011, Papadopoulos2013}. Interestingly, recent studies
also suggest that CRs might play an important role in galaxy
formation. CR diffusion leads to galactic-scale winds, which
effectively remove material from the disks. When included in detailed
3D hydrodynamical simulations, these CR-driven winds lead to more
realistic galaxy rotation curves \cite{Booth2013, Salem2014a}.

Although the impact of CRs on their environment is presumably very
significant on many spatial scales, their observational study is
rather challenging. The vast majority of CRs are charged atomic
nuclei, are deflected in interstellar and intergalactic magnetic
fields and lose directional information on their way to Earth. The
study of CR feedback hence requires indirect detection techniques,
including measurements \cite{Indriolo2013} of: 
\begin{enumerate}
\item abundances and abundance ratios of certain molecular ions,
\item X-ray line emission from electronic de-excitation, 
\item \g-ray line emission from nuclear de-excitation, 
\item observation of light-element isotope abundances, 
\item \g-ray emission from $\pi^0$-decay (protons), and 
\item \g-ray emission from inverse Compton (IC), synchrotron and
  bremsstrahlung processes (electrons). 
\end{enumerate}
In this review I will focus on the latter two channels and refer the
interested reader to the review by \citep{Indriolo2013} and references
therein for other possible observational signatures of e.g. CR
ionisation. Another emphasis will be put on the recent developments in
\g-ray astronomy in the study of starburst galaxies.  The interested
reader is referred to the more comprehensive review by
\citep{Bykov2014}, which also partly served as guidance for this work.

High-energy (HE; 100\,MeV $\leq E \leq$ 100\,GeV) and very high-energy
(VHE; 100\,GeV $\leq E \leq$ 100\,TeV) \g\ rays are tracers of
non-thermal processes of CRs with radiation fields, magnetic fields
and gas in the vicinity of particle accelerators. The vast majority of
Galactic particle accelerators observed at TeV energies is associated
with the end products of stellar evolution such as supernovae remnant
(SNR) shells, pulsar wind nebulae (PWNe) or \g-ray binary systems. The
\g-ray-emitting objects cluster tightly along the Galactic plane and
trace regions of dense gas and star formation. With their increased
SFR and hence supernova (SN) explosion rate, starburst galaxies are
ideal objects to study the physics of CRs and their impact on the ISM
and the overall galaxy dynamics with \g\ rays.

The non-thermal emission from starburst galaxies can be studied at
many wavelengths from low-frequency radio, X-rays up to GeV and TeV
\g\ rays. Radio observations probe low-energy electrons and dense gas,
and can be used to infer magnetic fields in starburst galaxies (see
e.g. \cite{Beck2011} and references therein for magnetic fields in
spiral and starburst galaxies). HE and VHE electrons can be probed via
synchrotron emission that is radiated at X-ray wavelength. The same
population of electrons up-scatters far-infrared photons, which
originate from the strong dust-reprocessed stellar radiation, to GeV
and TeV energies. X-ray observations in combination with \g-ray
measurements therefore provide powerful tools to probe the environment
in starburst galaxies. Electrons, however, do not comprise the
dominant component of CRs. Hadronic CRs are much harder to probe, as
e.g. synchrotron emission is suppressed by a factor
$(m_p/m_e)^4$. Energetic protons and heavier nuclei undergo
proton-proton interactions with dense gas particles and produce
neutral and charged mesons for proton energies above
$\sim$300\,MeV. The $\pi^0$'s instantly decay into two \g\ rays that
can be measured above energies of $\sim$70\,MeV. Charged pions decay
into electrons and positrons as well as neutrinos.

Observations of starburst galaxies at \g-ray energies provide a
multitude of information. They firstly tell us how efficient CRs
accelerated in SNRs and other particle accelerators are converted into
\g-ray emission in interactions with gas in the starburst
region. Secondly, the measured \g-ray luminosity and inferred CR
energy densities can be combined with measurements at lower energies
to study the ISM conditions in starburst regions. The third question
that can be answered with \g-ray observations is whether or not
equipartition between CRs, magnetic fields and radiation fields hold
in the extreme environments of starburst galaxies. With their greatly
enhanced SFR and SN rate, starburst galaxies offer an independent
probe for the SNR paradigm for CR origin.

\section{\g-ray emission from starburst galaxies -- Observations}
\label{sec:observation}

\subsection{GeV and TeV observations of starburst galaxies}
Given the expected CR energy input from SN explosions and the dense
gas present in starburst nuclei, the nearby starburst galaxies M82 and
NGC\,253 have long been predicted to emit \g\ rays at a detectable
level \cite[e.g.][]{Voelk1989, Akyuz1991, Paglione1996}. The previous
generation of satellite-based instruments such as EGRET, and
ground-based imaging atmospheric Cherenkov telescopes like HEGRA were,
however, not quite sensitive enough to detect this \g-ray emission and
only reported upper limits \cite{Blom1999, Goetting2007}.

It was the currently operating generation of \g-ray instruments that
finally allowed us to detect NGC\,253 and M82. The H.E.S.S. and
VERITAS collaborations reported on the detection of both objects in
VHE \g\ rays in 2009 \cite{Acero2009, Acciari2009}. Shortly after the
TeV discoveries, the Fermi collaboration reported on the detection of
both starbursts at GeV energies with the Large Area Telescope (LAT) in
2010 \cite{Abdo2010}. Since then, the Fermi-LAT GeV data sets
increased compared to the original publication \cite{Ackermann2012}
and H.E.S.S. published a detailed spectral and morphological study of
NGC\,253 \cite{Abramowski2012}.

\begin{figure}[!t]
  \centering
  \subfigure[]{\includegraphics[width=7.7cm]{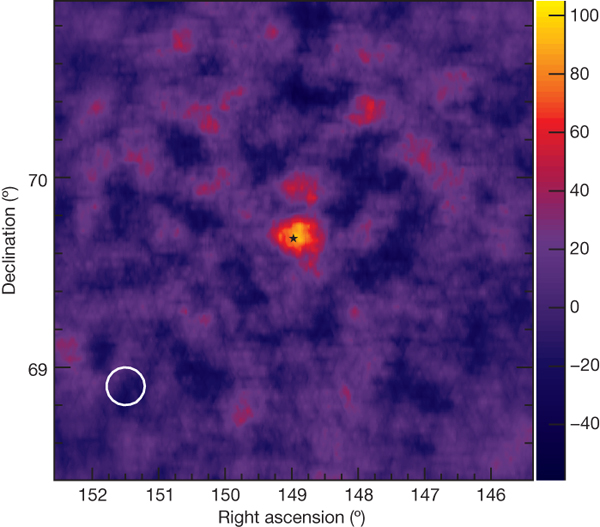}}
  \subfigure[]{\includegraphics[width=7.3cm]{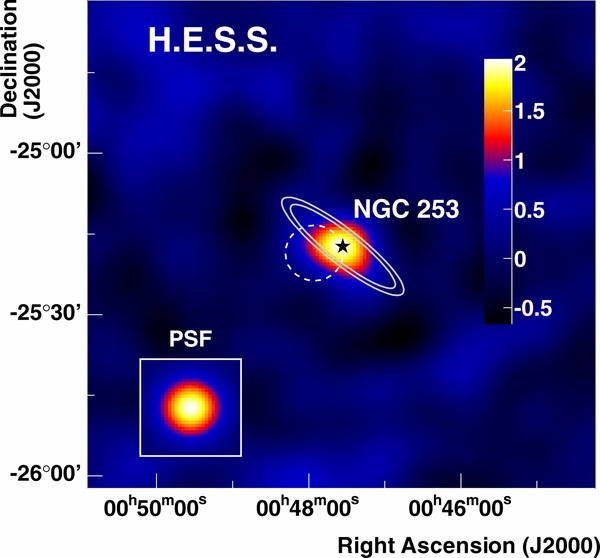}}
  \subfigure[]{\includegraphics[width=4.55cm]{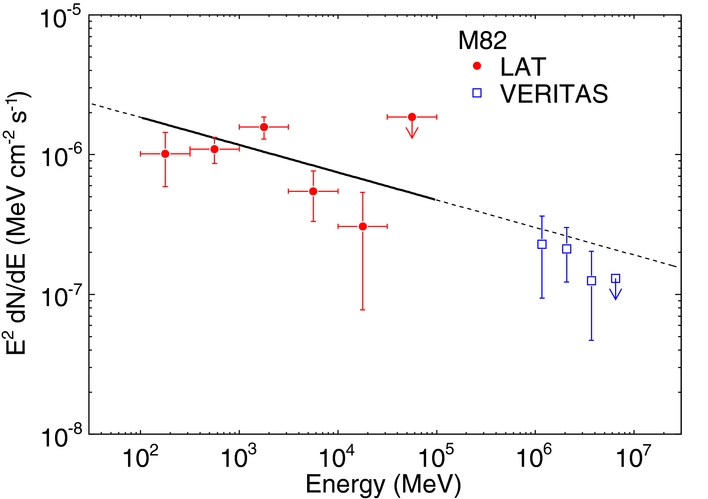}}
  \subfigure[]{\includegraphics[width=5.2cm]{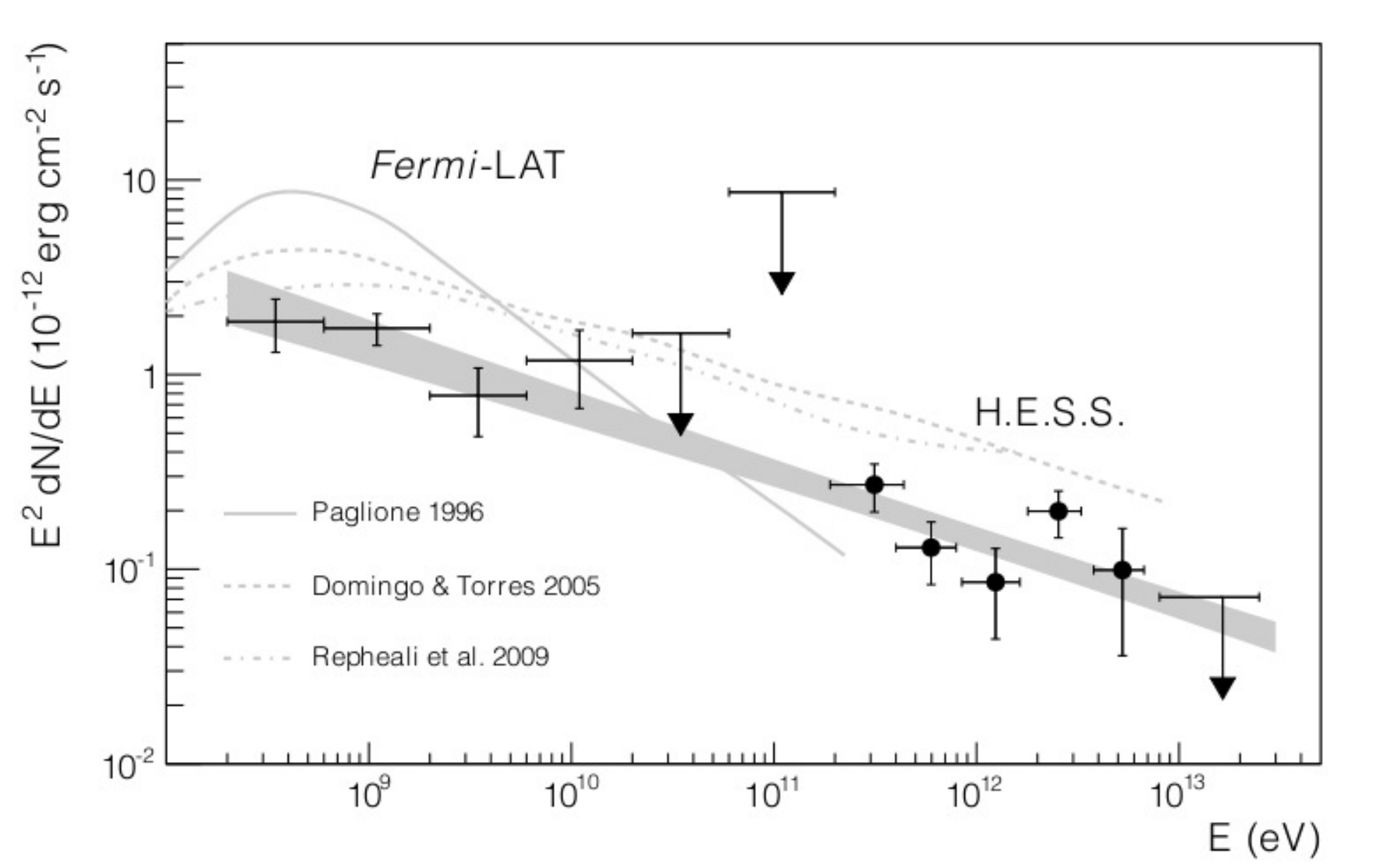}}
  \subfigure[]{\includegraphics[width=5.35cm]{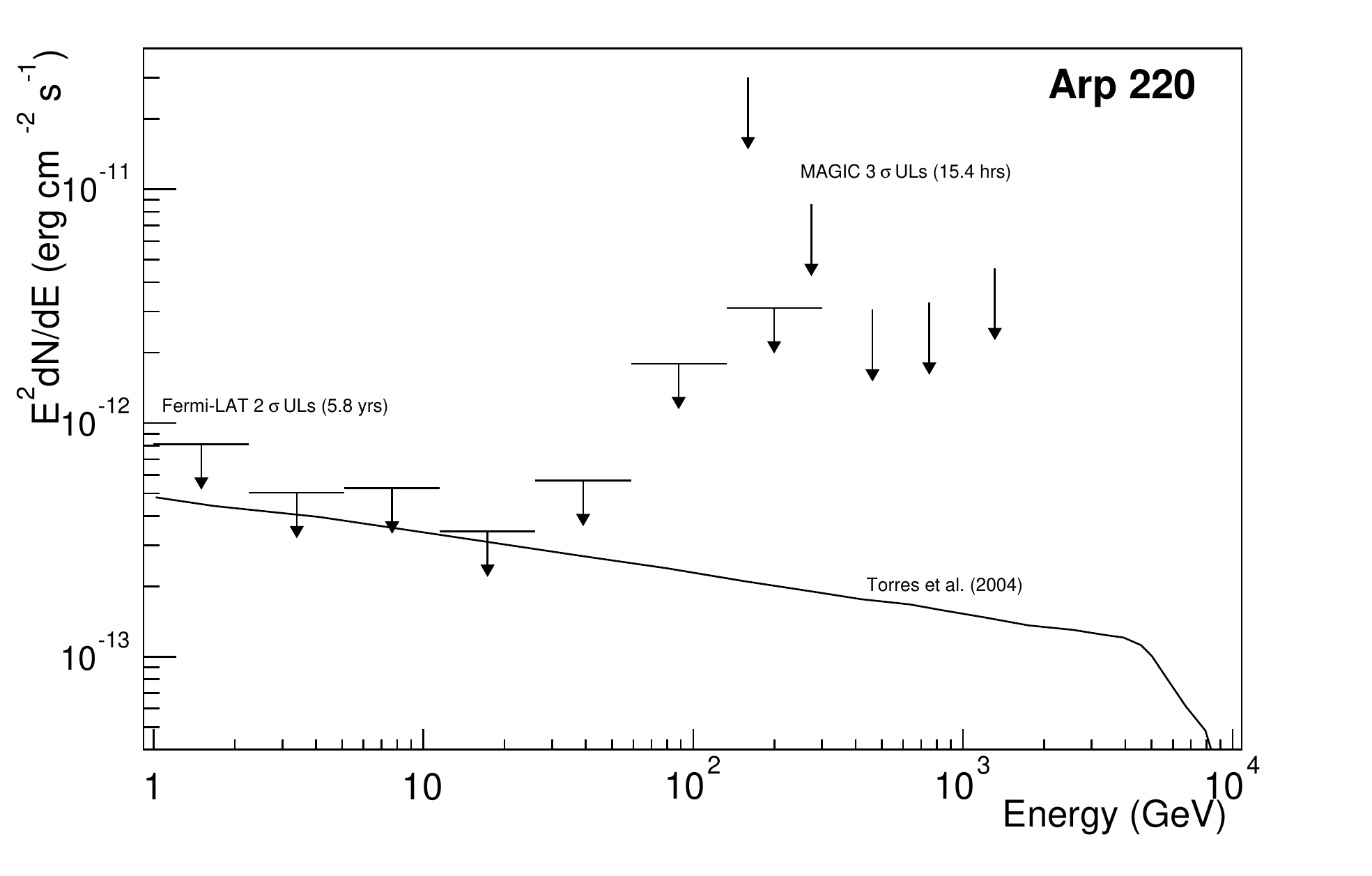}}
  \caption{Sky maps of M82 (a, \cite{Acciari2009}) and NGC\,253 (b,
    \cite{Abramowski2012}), and their combined GeV and TeV \g-ray
    spectra (c, \cite{Ackermann2012}; d, \cite{Abramowski2012}). e)
    Arp\,220: \g-ray flux upper limits from 5.8 years of Fermi-LAT
    data along with the MAGIC data and model predictions (full line,
    \cite{Albert2007}). Note the conversion of energy flux into SI
    units is 1\,W\,m$^{-2}$ =
    $10^{3}$\,erg\,cm$^{-2}$\,s$^{-1} = 6.24\times
    10^{8}$\,MeV\,cm$^{-2}$\,s$^{-1}$.}
  \label{fig:starbursts}
\end{figure}
Figure~\ref{fig:starbursts} shows the sky maps of M82 and NGC\,253 for
\g\ rays with energies $\gtrsim$200\,GeV and the \g-ray spectra over
almost five orders of magnitude in energy from $\sim$100\,MeV to
$\sim$10\,TeV. The \g-ray emission from both starbursts appears to be
point-like with a limit on the possible extension comparable to the
size of the central starburst in the case of NGC\,253. The \g-ray
emission is spatially coincident with the starburst nuclei of NGC\,253
and M82. On a qualitative basis, this observation supports the idea
that the regions of enhanced star formation and supernova activity are
also the regions where the bulk of CRs are being produced, interact
with the dense gas, and produce the observed \g-ray emission. The
\g-ray signals are also not time-variable, suggesting that a buried
central active galactic nucleus is not significantly contributing to
the \g-ray emission. The \g-ray spectra of M82 and NGC\,253 look
surprisingly similar and can both be described by a single power-law
in energy $F(E) = F_0 E^{-\Gamma}$, with $\Gamma \sim 2.2 \pm 0.1$.
Although statistics is limited, there is no indication of a spectral
break or cutoff feature apparent in the data, which suggests that
energy-independent transport and/or losses dominate in these
systems. The measured \g-ray flux level further suggests that (in the
most simplistic picture; see below) only a fraction of CRs produced in
SN explosions in these two starbursts interact and produce \g-ray
emission in proton-proton interactions. In the next section I will
briefly review the interpretation of these results and what they tell
us about the properties of non-thermal particles in starburst galaxies
and their impact on the ISM. But let us briefly look at other galaxies
that can potentially be detected at \g-ray energies.

\subsection{GeV and TeV observations of other star-forming galaxies}
The link between star formation and CR acceleration and subsequent
\g-ray production is not unique to starburst galaxies or the Milky Way
but applies to any star-forming galaxy. The Fermi-LAT collaboration
searched for GeV \g-ray emission from a sample of 69 nearby
star-forming galaxies (dwarfs, spirals, LIRGs and ULIRGs) and studied
the connection with star-formation tracers such as radio and
far-infrared emission over five orders of magnitude in SFR
\cite{Ackermann2012}. Apart from \g-ray emission from M82 and
NGC\,253, the LAT detected \textit{diffuse} GeV emission from the
Milky Way, its satellite galaxies the Large and Small Magellanic
Clouds (LMC, SMC), and the Andromeda galaxy (M31). GeV \g\ rays were
also observed from the LIRG NGC\,2146 \cite{Tang2014} and the
Seyfert-II galaxies NGC\,1068 and NGC\,4945 \cite{Lenain2010}. For the
latter two, however, the contribution associated with star-forming
activity is hard to disentangle from the central (active) black
hole. Even with almost six years of Fermi-LAT data, the most active
star-forming galaxy in the local universe, the ULIRG Arp\,220 is not
detected in GeV \g\ rays. The derived upper limits are close to
theoretical model predictions and start to probe the fraction of CRs
that interact in Arp\,220's starburst region (see
Fig.~\ref{fig:starbursts}).

At TeV energies, H.E.S.S. reported the detection of diffuse emission
from the Milky Way \cite{Abramowski2014} and discovered the first
population of particle accelerators at TeV energies in an external
galaxy: the LMC \cite{Abramowski2015}. Among the discovered sources
are the PWN associated with the most energetic pulsar known, the
prominent SNR N\,132D and the largest non-thermal X-ray shell, the
30\,Dor\,C superbubble -- the first of its kind seen at TeV
energies. However, no diffuse emission from CRs interacting with the
ISM were detected in these observations. For Andromeda, the VERITAS
collaboration reported upper limits on the TeV \g-ray emission using a
limited data set and not-yet optimised/final analysis \cite{Bird2014}.
All these discoveries have been made over the past five years,
demonstrating that current \g-ray instrument have now reached the
sensitivity to probe CRs in external galaxies. In the following I will
briefly discuss the interpretation of the \g-ray results and give a
short overview of current theoretical work.

\section{\g-ray emission from starburst galaxies -- Interpretation}
\label{sec:interpretation}

\subsection{The physical conditions in starburst regions}
\label{sec:ISM}

In order to study the production, propagation and interaction of
non-thermal particles in starbursts, it is important to look at the
physical conditions in star-forming regions. The ISM in a starburst
region differs significantly from the conditions in the typical Milky
Way ISM. Gas densities in starbursts are much higher than in the
Galactic ISM and the rate at which new stars form is greatly
enhanced. The large number of massive stars leads to a high number
density of stellar (mainly UV) photons that are absorbed by dust,
which re-emits at infrared wavelength. Magnetic fields are also
enhanced by a similar factor. As a result, non-thermal particles
experience stronger losses and cool faster. The dense gas in the
starburst region is heated by stellar winds and SNe and the additional
energy input can not be compensated for by radiative cooling. As a
result, a starburst wind (or ``superwind'') develops, which
effectively removes gas and non-thermal particles from the core and
enriches the galactic halo. At the same time, the role of particle
diffusion relative to advection is diminished due to nonlinear effects
caused by the increased energy flux density of CRs and the magnetic
field fluctuations they drive (see e.g. \cite{Abramowski2012}). For
the typical dimensions of starburst regions in M82 or NGC\,253 for
instance, the advection time in the superwind is $\sim10^5$ years ---
much shorter than the starburst lifetime of $\gtrsim10^7$ years. This
implies that accelerated particles spend only a limited time in the
starburst region and that the non-thermal particle spectrum reaches an
equilibrium after a few $10^5$ years.

\begin{figure}
  \centering
  \subfigure{\includegraphics[width=7.3cm]{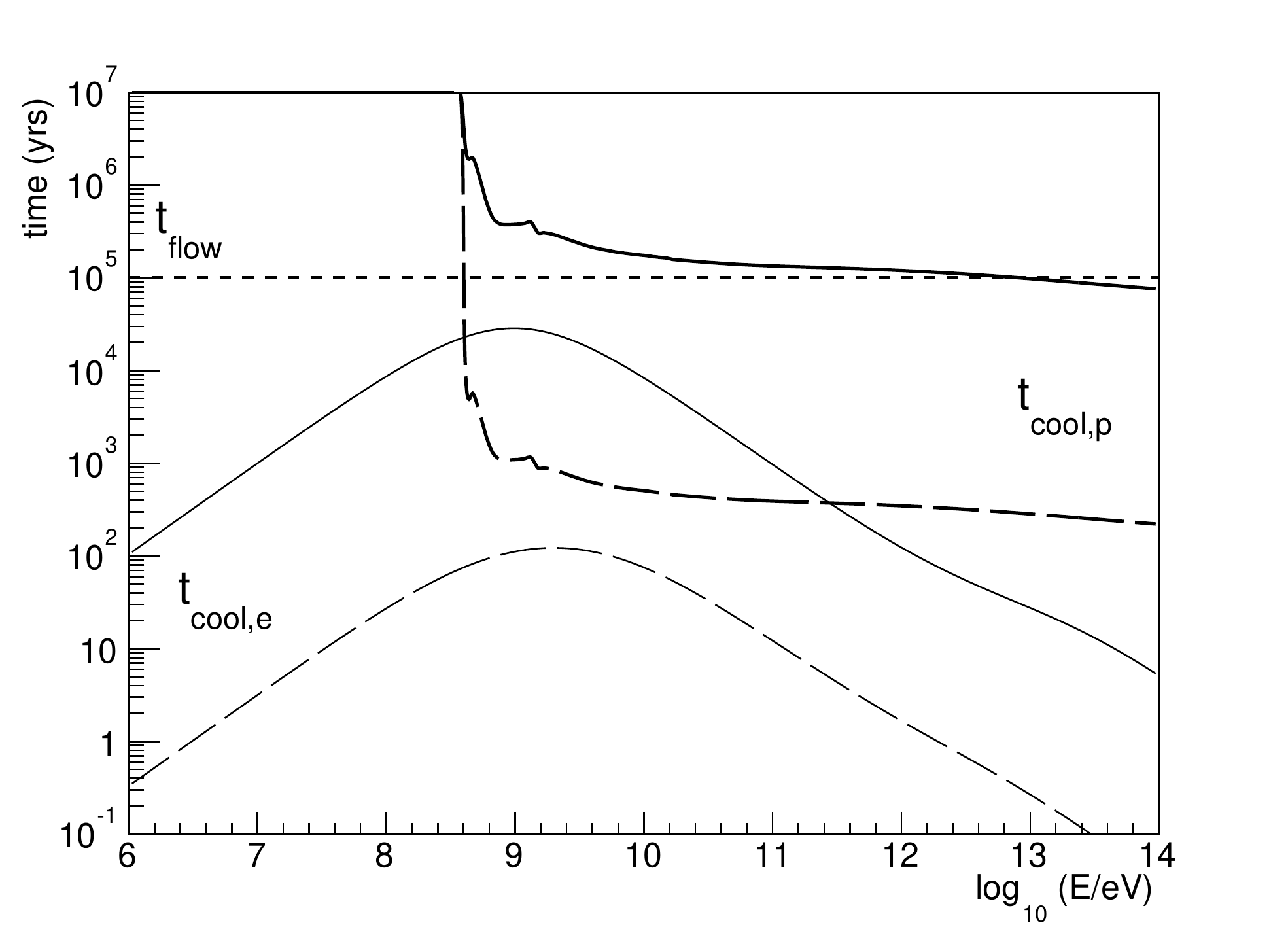}}
  \subfigure{\includegraphics[width=7.7cm]{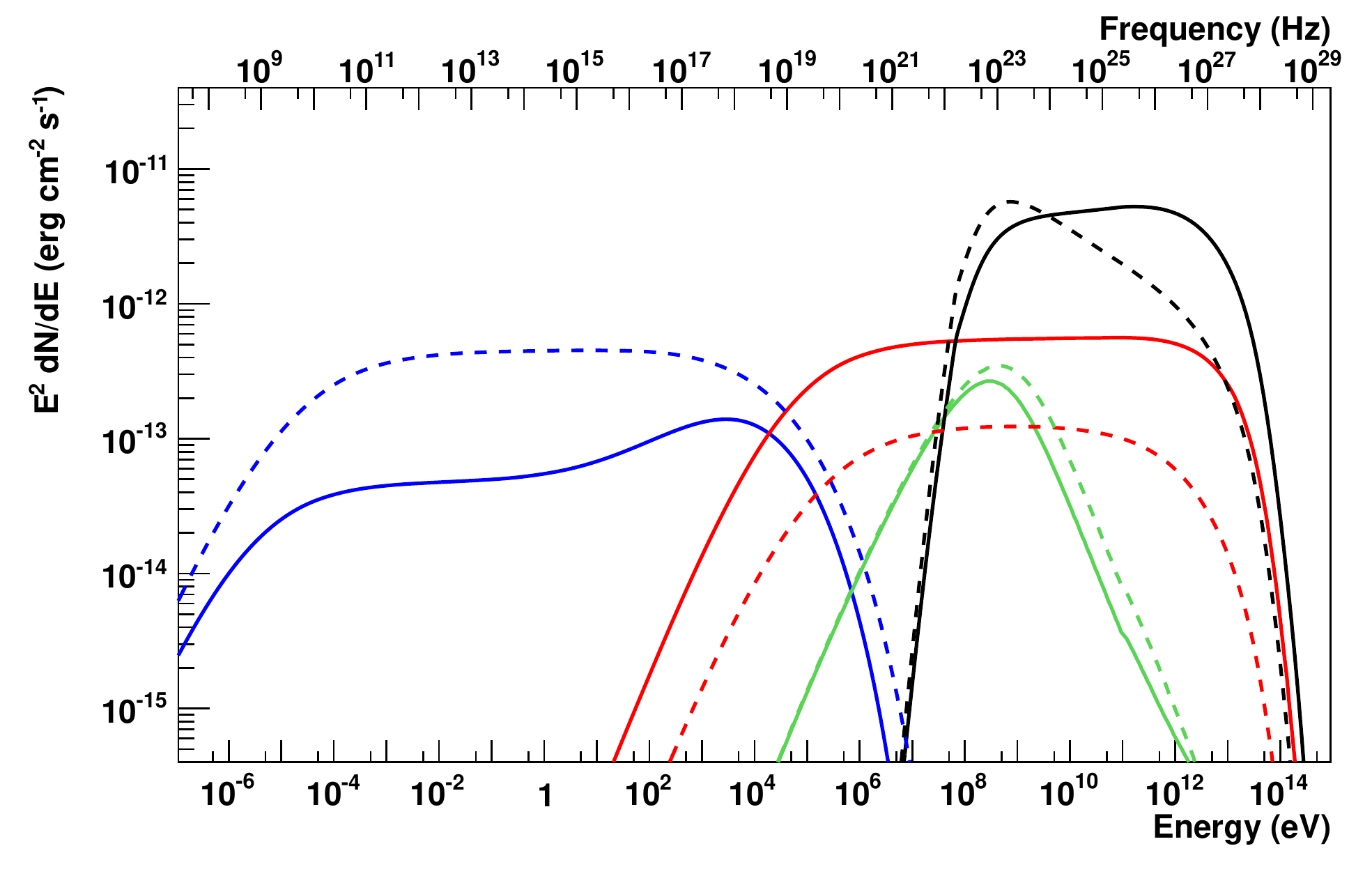}}
  \caption{(a) Relevant cooling timescales for electrons and protons
    in a starburst galaxy such as M82 or NGC\,253 (full lines) and for
    a ULIRG such as Arp\,220 (long dashed lines). The short-dashed
    line indicates a flow timescale of the central starburst winds of
    100\,kyr. Electron cooling considers Coulomb, bremsstrahlung,
    synchrotron and IC processes (taking into account the full
    Klein-Nishina cross section). Physical parameters used for the
    timescale calculation are summarised in Table~\ref{tab:model}. (b)
    Model spectral energy distributions for two representative
    starburst galaxies and contributions from IC (red), synchrotron
    (blue), bremsstrahlung (green) and $\pi^0$-decay \g\ rays
    (black). Full lines represent the low-magnetic, high-radiation
    field energy density case. Dashed lines depict the high-magnetic,
    low-radiation energy density case. The index of injected protons
    is 2.0 (full), 2.3 (dashed), and 2.0 for electrons in both
    scenarios \cite[Figure taken from][]{Ohm2012}.}
  \label{fig:cool}
\end{figure}

\begin{table}
  \centering
  \begin{tabular}{lcccccc}
    \hline
    \hline
    Object & distance & n$_{\rm{H}}$ & $B$ & $T_{\rm{IR}}$ &
    $u_{\rm{IR}}$ & References\\
           & Mpc & cm$^{-3}$ & nT & K & eV\,cm$^{-3}$ & \\ \hline
    NGC\,253 & 3.4 & 580 & 15 & 50 & 2400 & 
    \citet{Abramowski2012, Heesen2011, Ohm2013}\\
    Arp\,220 & 74 & $\gtrsim$10$^5$ & 250 & 100 & $\sim$$1.5\times10^5$ &
    \citet{Scoville2014, McBride2015}\\
    \hline
    \hline
  \end{tabular}
  \caption{Physical quantities, typical for the starburst
    environment of NGC\,253 the ULIRG Arp\,220.}
  \label{tab:model} 
\end{table}

Figure~\ref{fig:cool}a) shows the cooling times of electrons and
protons computed for the starburst nuclei of NGC\,253 and
Arp\,220. Also shown is the typical advection timescale of the gas to
leave the starburst region in the central molecular outflow. In a
typical starburst galaxy like NGC\,253, the advection timescale is
comparable to the proton-proton cooling time, but much longer than the
electron cooling time. This implies that electrons predominantly stay
and cool in the starburst environment, whereas protons either interact
with the dense gas or leave in the superwind. The situation in
Arp\,220 is even more extreme, as basically \emph{all} accelerated
protons and electrons cool in the starburst region. The cooling time
of TeV electrons is as short as one year (due to efficient synchrotron
and IC cooling) and still only $\sim$100
years at 1\,GeV energies, where bremsstrahlung dominates. At electron
energies below 1\,GeV, Coulomb losses become increasingly important
and limit the cooling time again to less than 100 years. These short
cooling timescales are important when considering the sources of
non-thermal particles and comparing them to HE and VHE \g-ray sources
and source populations in the Milky Way, as discussed in the next
section.

\subsection{Origin of the \g-ray emission}
First model predictions for the non-thermal emission from starburst
galaxies were made more than 20 years ago \cite{Voelk1989,
  Paglione1996} and predicted source fluxes at the sensitivity limit
of running or planned instruments at that time. Early model
predictions and most of the following more detailed calculations are
one-zone models and assume the \g-ray emission to be of diffuse origin
\cite[e.g.][]{Domingo2005, Rephaeli2010, Lacki2010, Lacki2011,
  Paglione2012, Abramowski2012, Yoast-Hull2013}. Depending on the
complexity of the study, these models consider numerous non-thermal
processes of relativistic electrons and protons as outlined above,
such as:
\begin{enumerate}
\item neutral and charged pion production by protons and heavier
  nuclei, as well as subsequent pion-decay products;
\item synchrotron, IC, bremsstrahlung and Coulomb losses of energetic
  electrons/positrons;
\item diffusive escape of CRs and advection in the starburst wind. 
\end{enumerate}
All of these models assume that SNe are the source of non-thermal
particles, that a constant fraction of kinetic energy per SN explosion
is transferred into CRs, and that these CRs then propagate, interact
in the starburst region and finally leave the system in the starburst
wind. The predicted emission from electrons and hadrons is then
compared to measurements at radio wavelengths and \g-ray energies. The
predicted radiation spectra are sensitive to the physical conditions
in the starburst region such as the magnetic field, gas density (and
structure), and radiation fields. An important ingredient in the
modelling is the electron-to-proton (e/p) ratio, which is inferred
from Galactic radio measurements and under the assumption of
equipartition between magnetic fields and CRs, and is typically found
to be 1/100 in the Milky Way.

To illustrate the influence of different quantities on the radiation
spectra, Figure~\ref{fig:cool}b) shows two simple one-zone,
time-dependent models for the continuous injection of electrons and
protons over 200\,kyrs in two representative starburst environments at
3.5\,Mpc (see \cite{Ohm2012} for a detailed model description). The
first model shows the SED for a starburst galaxy where IC losses in
the strong radiation fields dominate over synchrotron emission (full
lines). The second model illustrates what happens if the magnetic
field is enhanced by a factor of two and the radiation field energy
density lowered by an order of magnitude. In this case, the
synchrotron component at X-rays is much brighter than the IC component
seen in \g\ rays. For protons, two different injection spectra are
assumed to illustrate the effect on the predicted \g-ray
emission. Changing the radiation field energy densities and magnetic
fields in the two cases does not affect the $\pi^0$-decay \g-ray
spectrum. From Figure~\ref{fig:cool}b) it is also clear that even for
the assumed large e/p ratio of 1/10, hadrons dominate the \g-ray
emission at almost all energies. Only at energies close to the $\pi^0$
production threshold and at multi-TeV energies, electrons can
significantly contribute to the emission in this simplified scenario,
and for the rather extreme choice of the e/p ratio. This model (and
all studies mentioned before) suggest that the diffuse emission in
starburst galaxies is of \textit{hadronic origin}. As inelastic
proton-proton collisions produce not only neutral, but also charged
pions, which further decay into electrons and positrons, these
secondary leptons also cool in the starburst ISM. \citet{Lacki2013}
have shown that secondaries can significantly contribute, if not
dominate, the radio emission in starburst environments. Together with
the \g-ray measurement, this gives us an independent measure of the
equipartition magnetic field in starburst galaxies and galaxies, where
hadrons dominate the \g-ray emission.

\subsection{Proton calorimetry and individual source populations}
\label{sec:cal}

As briefly discussed in Section~\ref{sec:ISM} CR diffusion and escape
in the starburst wind are the two processes that compete with the
\g-ray production in a hadronic scenario, and determine the level of
emission seen at \g-ray energies. Following the H.E.S.S. collaboration
(\cite{Abramowski2012} and references therein) the energy in
interacting CRs $L_{\rm coll}$ that leads to the measured \g-ray
emission level
\begin{equation}\label{eq1}
L_{\rm coll} = 4 \pi d^2 F_{\gamma}^{\rm meas}
\end{equation} 
can be compared to the total energy in CRs $L_{\rm CR}$ that is in
principle available for \g-ray production
\begin{equation}\label{eq2}
L_{\rm CR}(\pi) = f_{\pi}\nu_{\rm SN}\epsilon E_{\rm SN}. 
\end{equation}
The main uncertainties in these quantities come from the estimates of
the distance $d$, and the product of SN kinetic energy $E_{\rm SN}$
and conversion efficiency into CRs $\epsilon$. $f_\pi$ is a correction
factor, accounting for the \g-ray spectral index and underlying
available power per CR energy interval ($f_\pi = 1$ for a \g-ray
spectral index of 2.0) and $\nu_{\rm SN}$ is the SN rate. The ratio
$L_{\rm coll} / L_{\rm CR}(\pi)$ is often referred to as the
\emph{calorimetric fraction} and its value tells us how efficient a
starburst galaxy (or a star-forming galaxy in general) is in
converting CRs into hadronic \g-ray emission. The values found for M82
and NGC\,253 are of the order of $20\% - 40\%$ \cite{Lacki2011,
  Abramowski2012} and agree well with the rough comparison of
proton-proton cooling time and escape time in Fig.~\ref{fig:cool}. The
CR energy density of pion-producing particles in NGC\,253's starburst
region is a few hundred times larger than in the Milky Way. The
inferred magnetic field strength, assuming equipartition, is
$\sim$10\,nT \cite{Abramowski2012} and consistent with estimates by
\citet{Heesen2011}.

All models discussed so far consider SNRs as the sole source of
high-energy particles in starburst galaxies. The population of
Galactic GeV and TeV \g-ray sources, however, is diverse and dominated
by PWNe, rather than SNRs. \citet{Mannheim2012} proposed PWNe as the
source class, responsible for the bulk of the TeV emission from
NGC\,253 and M82. In their work, they scaled the population of Milky
Way PWNe to both starbursts. As discussed in Section~\ref{sec:ISM} the
physical conditions in starburst galaxies, however, are significantly
different from the typical ISM in the Galaxy and cooling times for
TeV-emitting electrons are very short. \citet{Ohm2013} performed a
more detailed modelling of the PWN populations in NGC\,253 and M82,
using a time-dependent two-zone model. They consider a stage where
particles are accelerated and cool in the PWN environment early on,
and a later stage where particles escape the PWN and cool in the
starburst region. Although assumptions are very different,
\cite{Mannheim2012} and \cite{Ohm2013} reach similar conclusions,
namely that PWNe are potential contributors to the TeV emission from
starburst galaxies. Note that for the latter model, a steeper spectral
index in the TeV range requires pulsars to have very short birth
periods of $\sim$15\,ms and fast particle escape in order not to
violate radio measurements. Both models predict a feature, where
(energy-dependent) diffusion results in a steepening of the \g-ray
spectrum, and where the harder \g-ray spectrum produced by PWNe starts
to dominate the emission. The sensitivity of current instruments is
not yet good enough to probe such a feature, but it should be noted
that no indication of a break or cutoff in the NGC\,253 spectrum is
apparent, which would support the PWN interpretation.

\section{Starburst galaxies in the CTA era}
\label{sec:cta}

As is clear from the previous sections, starburst galaxies are weak
\g-ray emitters and only a limited number of nearby objects can be
studied. Any improvement in statistics, which results in better
measured \g-ray spectra requires either longer observations and/or
better \g-ray detectors. Fermi-LAT continues to collect data of
starburst galaxies thanks to its all-sky capabilities, and will
improve the used analysis techniques. This will provide a better
determination of the \g-ray spectrum at GeV energies. The angular
resolution might not be sufficient to resolve the starburst nucleus,
or to even detect an intrinsic extension of the \g-ray source. For
Cherenkov telescopes, on the other hand, increasing the exposure does
not improve things too much, since these instruments are already
operating at the sensitivity limit. This is especially true for
NGC\,253, were already more than 150\,hours of data have been
collected and for which advanced analysis techniques are
used\footnote{A better description of the M82 TeV spectrum may be
  possible by employing advanced analysis techniques and by increasing
  the observation time with VERITAS. MAGIC observations could improve
  the spectral coverage below $\sim 500$ GeV.}.

\begin{figure}[!t]
  \centering
  \includegraphics[width=11.5cm]{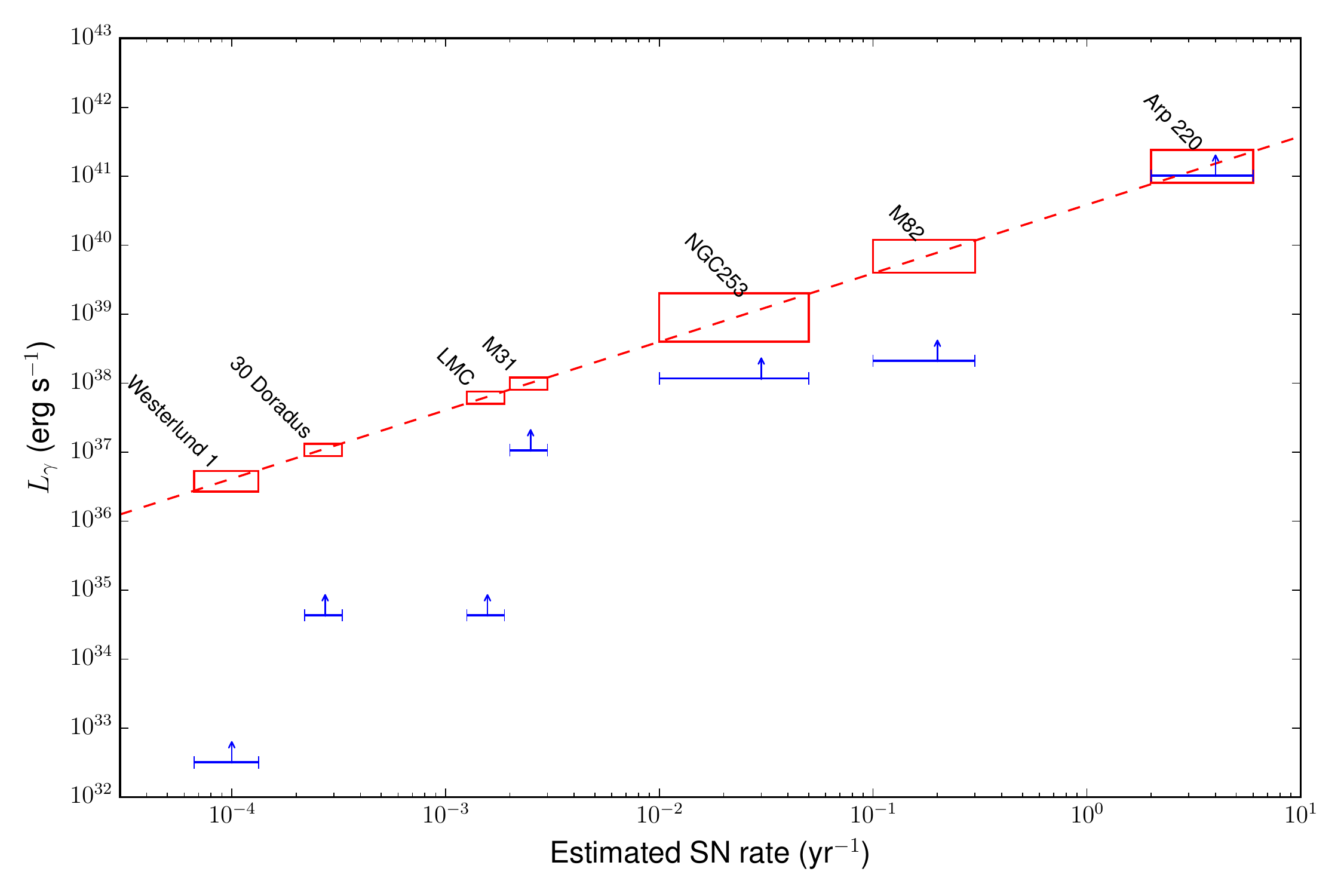}
  \caption{Expected calorimetric \g-ray flux of nearby star-forming
    galaxies with varying star-formation and SN rates. Red boxes
    indicate uncertainties in SN rates and distance of the source. See
    text for a detailed description. The expected CTA sensitivity is
    shown as blue arrows and the red dashed line indicates the
    expected \g-ray flux in the calorimetric limit (see text).}
  \label{fig:corr}
\end{figure}

The sensitivity of Fermi-LAT and current ground-based Cherenkov
telescopes is not sufficient to search for spectral signatures
associated with different source populations, different energy-loss
processes or \g-\g\ absorption features. A significant improvement in
sensitivity for energies between several tens of GeV and hundreds of
TeV energies is expected from the upcoming Cherenkov Telescope Array
(CTA) \cite{Actis2011}. CTA is currently in the prototyping phase,
will consist of order 100 telescopes and be located in two sites, one
in the northern and the other in the southern hemisphere (see
\cite{Knoedlseder15} in this issue for more details on the future of
\g-ray astronomy). In the following I will outline the CTA potential
for the study of starburst galaxies, the measurement of the
calorimetric fraction and how to extend the sample to nearby
star-forming galaxies and Galactic star-forming regions -- similar to
what has been done by the Fermi-LAT collaboration \cite{Ackermann2012}
at GeV energies. Here we follow the approach described in
\cite{Abramowski2012} and calculate the \g-ray flux above 300\,GeV in
the calorimetric limit where $L_{\rm coll} = L_{\rm CR}(\pi)$ using
Eq.~2. This estimate mainly depends on the SN rate of the galaxy, the
energy in CRs per SN explosion, and the assumed \g-ray spectral
index. This flux is then compared to the sensitivity of CTA. Rather
than making a detailed model for each individual source using the
expected CTA sensitivities as presented in \cite{Bernlohr2013}, we
simply assume a ten times better sensitivity than
H.E.S.S. \cite{Aharonian2006} and a 100-hour observation. This results
in an energy flux sensitivity above 300\,GeV for an assumed spectral
index of $\Gamma=2.2$ of $\sim$$1.5\times10^{-16}$\,W\,m$^{-2}$
(or 0.1\% Crab). For several reasons this is only a rough guidance:
more advanced analysis techniques will be employed; northern and
southern arrays will have different sensitivities; and observation
times and conditions will most likely vary from one source to
another. Figure~\ref{fig:corr} shows the expected calorimetric flux
from stellar clusters in the Milky Way and the LMC, the nearby
star-forming galaxies M31 and the LMC, the starburst galaxies
NGC\,253, and M82, and the ULIRG Arp\,220. Also shown is the assumed
CTA sensitivity. First of all, this simple model already shows that
CTA is very well suited to study the connection between \g\ rays and
the star-formation process over a wide range of SFR and SN rates. It
is also clear that even for a long exposure, the predicted flux from
the ULIRG Arp\,220 is at the limit of the CTA sensitivity. Another
limitation of this model is that it assumes an equilibrium
situation. While this assumption is certainly valid for entire
galaxies, for which the SN rate is shorter than the typical
acceleration time in SNRs, this is not true for objects like
Westerlund~1, or 30~Doradus. Here the measured \g-ray emission level
may depend on the recent SN explosion history.

It is also important to perform detailed spectral and morphological
studies of individual objects with CTA. It will for instance be
possible to probe the \g-\g\ absorption feature, predicted to lead to
a cutoff in the spectrum at several TeVs \cite{Acero2013}. Although
challenging, the CTA angular resolution might be sufficient to detect
an extension of the \g-ray source and to confirm the origin in the
starburst nucleus. The \g-ray emission from the disk of NGC\,253 is
expected to be an order of magnitude dimmer than the starburst nucleus
\cite{Abramowski2012}. However,the CTA sensitivity will be an order of
magnitude better than H.E.S.S., which should allow us to constrain the
emission from the disk component. While the next major observational
progress is expected soon, the modelling of the \g-ray emission from
starburst galaxies will likely extend existing one-zone models. Some
authors for instance have already started to consider the multiple gas
components in the starburst nucleus, while others have extended
single-zone models to multiple zones, or considered different
non-thermal source populations in the extreme environments of
starbursts.

Starburst galaxies are in many respects interesting objects: 
\begin{enumerate}
\item their study is important for the understanding of galaxy formation and
evolution;
\item their measured \g-ray emission and the \g-ray emission from
  star-forming galaxies in general can comprise up to a quarter of the
  extragalactic background light \cite{Ackermann2012};
\item some significant fraction of the astrophysical neutrino flux,
  first detected with IceCube \cite{Aartsen2014}, can possibly be
  explained by star-forming galaxies for rather hard source spectra as
  observed at \g-ray energies \cite{Anchordoqui2014, Tamborra2014,
    Chang2014};
\end{enumerate}
Another possibility is that CRs, pre-accelerated in the starburst
nucleus can potentially be accelerated beyond the knee of the CR
spectrum in the galactic wind \cite{Dorfi2012}. To summarise,
observations of star-forming galaxies in general and starbursts in
particular have made major progress in the past five years with new
and exciting discoveries. There is ongoing development in the
theoretical modelling of CRs and their impact on the star-formation
process. With the advent of CTA and other major facilities at lower
wavelengths, exciting times are ahead of us for the study of the
connection of CRs and the star-formation process.

\section*{Acknowledgements}
I'd like to warmly thank G\'erard Fontaine and Bernard Degrange for
the invitation to write this review, Heinz V\"olk for his constant
interest and support, as well as Matthias F\"ussling for carefully
reading the manuscript.
\bibliographystyle{unsrtnat}
\bibliography{cras-starbursts}		


\end{document}